\documentclass{aastex}
\begin{document}
\def\etal{{\it et al.\/}}
\def\cf{{\it cf.\/}}
\def\ie{{\it i.e.\/}}
\def\eg{{\it e.g.\/}}

\title{SupraNova Events from Spun--up Neutron Stars:
an Explosion in Search of an Observation}
\author{{\bf Mario Vietri$^1$ and Luigi Stella$^2$}}
\affil{
$^1$Universit\`a di Roma 3, Via della Vasca Navale 84, 00147 Roma, \\
Italy, E-mail: vietri@corelli.fis.uniroma3.it \\
$^2$ Osservatorio Astronomico di Roma, Via Frascati 33,
00040 Monteporzio Catone
(Roma)\\
Italy, E-mail: stella@heads.mporzio.astro.it\\
Affiliated to I.C.R.A.\\
}

\begin{abstract}
We consider a formation scenario for supramassive neutron stars (SMNSs)
taking place through mass and angular momentum transfer from a close
companion during a Low Mass X--ray Binary (LMXB) phase, with the ensuing
suppression of the magnetic field. We show that this formation channel is
likely to work for all equations of state except the stiffest ones, 
L, N$^\star$. After the end of the mass transfer phase,
SMNSs will loose through magnetic dipole radiation most of their
angular momentum, triggering the star's collapse to a black hole. We
discuss the rate of occurrence of these collapses, and propose that these
stars, because of the baryon--clear environment in which the
implosion/explosion takes place, are the originators of gamma ray bursts.
\end{abstract}


\section{Introduction}

One of the key requests on gamma ray burst (GRB) models is that they make
contact with the fireball model (Rees and M\`esz\`aros 1992) which has
proven so successful in predicting and interpreting the observed
properties of GRBs' afterglows. In particular, this entails that a large
explosion is to take place in a region with small baryon contamination: 
for $E = 10^{53}$~erg, the baryon contamination must be
at most $E/\gamma c^2 \approx 10^{-4}$~M$_\odot$, for $\gamma = 300$, the bulk
Lorenz factor of the explosion. Vietri and Stella (1998) 
presented a model which could accomplish this, involving a supramassive
neutron star (SMNS), \ie, a neutron star with a larger baryon number than
any normal neutron star because it derives part of its support against
self--gravity from the centrifugal force; these
supramassive stars cannot be slowed down to zero spin rate, because
they are so massive that, as they lose angular momentum, they become
unstable to black hole formation before reaching zero spin rate (Cook,
Shapiro and Teukolsky 1994a,b, Salgado \etal, 1994). The model consists of
the implosion/explosion (I/E) of a supramassive neutron star which has lost
through magnetic dipole radiation so much angular momentum that it must then
collapse to a black hole; the rotational energy of the small amount of
equatorial mass left behind because already in near centrifugal equilibrium
provides the energy source that powers the burst. 
%In this scenario, the trigger is not a dynamical event, which would entail 
%large baryon pollution,
%but an instability leading to the implosion of an object after a long latency 
%period which is used to evacuate the surroundings of baryons.

In paper I we proposed that SMNSs are formed in the SN explosion
of a core with too much mass and angular momentum to end up in a normal neutron
star. Though still nothing stands against this possibility, and we are
not reneging it, we have now realized that a different channel exists: mass
and angular momentum accretion from a companion in a Low Mass X--ray Binary
(LMXB). The following discussion is relevant to the formation of MilliSecond 
Pulsars (MSPs), and will mimick arguments used in discussing the evolution of 
LMXBs into MSPs. We first discuss how the accretion of large 
amounts of mass and angular momentum may be realized in Nature and then 
apply this scenario to GRBs.

\section{Mass and angular momentum accretion}

The main obstacle to accretion onto a normal neutron star of large amounts of
angular momentum from a close companion via an accretion disk is the neutron
star's magnetic field: the neutron star can rotate only so fast as to make 
the corotation and Alfv\`en
radii coincide, lest a propeller phase sets in, which
would actually entail angular momentum loss (Ghosh and Lamb 1978,
Illarionov and Sunyaev 1975). The coincidence of these two radii leads to an
equilibrium period, $P_{eq} = 1.3 (B/10^{12}\;G)^{6/7}
(\dot{M}/\dot{M}_{Edd})^{-3/7}$~s (Ghosh and Lamb 1992), which clearly shows
that $B$ must decrease before significant spin--up can occur.
Though no unique model has emerged yet, the current consensus
is that the neutron star magnetic field decays by at least $3-4$ decades 
either as a direct result of mass accretion (Phinney and Kulkarni 1994) 
or of the ensuing spinup (Ruderman, Zhu and Chen 1998).

The strongest constraints on field decay in NSs come from LMXBs and MSPs.
In only one LMXB, SAX~J1808.4-3658 a coherent 2.5~ms
signal has been detected in the persistent X-ray emission,
providing direct evidence for the presence
of a small magnetosphere; the inferred magnetic field is in the
$B\sim 10^8-10^9$~G range (Psaltis and Chakrabarty 1998).
All other LMXBs have undetectably small coherent pulsations in their
persistent emission, if at all. Yet spin periods in the in the 2-4~ms range 
have been deduced for about ten LMXBs from the X-ray flux
oscillations that are present during type I bursts emitted by these sources
(cf. van der Klis 1998). Spinup through accretion can have occurred in
these neutron stars only if their magnetic field is lower than
$\sim 10^9$~G. Further evidence that the field might have
decayed to dynamical insignificance derives from the modelling of the
kHz Quasi Periodic Oscillations (QPO), a common phenomenon observed
in LMXBs. In the sonic point model (Miller, Lamb and Psaltis 1998),
the Alfv\'en radius is located at a radius corresponding to a Keplerian
frequency of $\approx 350\; Hz$, corresponding to a magnetic field of
$\approx 8\times 10^8\; G$: this already bears witness to a
thousand--fold reduction of the magnetic field below that of a typical
newborn pulsar. A better model explains QPOs in terms of the fundamental 
frequencies of test particle motions in the general--relativistic potential 
well in the vicinity of the neutron star
(Stella and Vietri 1999). The model is capable of explaining the observed
relation between peak QPO frequency and its lower frequency counterpart over
three orders of magnitude in peak QPO frequency and several distinct classes
of sources, including candidate black holes and LMXBs (Stella, Vietri and
Morsink 1999), implying that the magnetic field has
been reduced already to $\la 2\times 10^8$~G during the LMXB phase.

A different argument involves a handful of MSPs with observed magnetic fields
$\la 2\times 10^8$~G , of which there are currently about a dozen, including
the lowest fields ever measured, $7\times 10^7\;G$ in $J2229+2643$
and $J2317+1439$ (Camilo, Nice and Taylor 1996). 
For these small fields, the Alfv\'en radius for disk accretion 
(Ghosh and Lamb 1978, 1992) is smaller than
$\sim 12$~km, which, according to Cook, Shapiro and Teukolsky (1994b)
is larger than the radius of the innermost stable orbit for the softest
equations of state for a neutron star with $M = 1.4$~M$_\odot$ (see Table 1). It
should also be noticed that it has been argued cogently (Arons 1993) that the
magnetic field in these objects is dominated by the dipole component, with
negligible contribution from higher multipoles. Altogether, this means that we
have already observed the result of accretion from a companion pushing the
magnetic field to dynamical irrelevance (or, at least, very close to it)
sometime during the mass exchange process. The argument about QPOs (Stella
and Vietri 1999) implies that this may happen reasonably early in the LMXB
history.

Detailed models are required to establish the exact history of a neutron
star's mass, angular momentum and magnetic field, but unfortunately these 
computations are currently fraught with uncertainties: where is the $B$ field 
located, in the core or in the crust? And what is an appropriate model for the 
field suffocation? Population synthetic studies of this phenomenon (Possenti
\etal, 1999) have focused on two representative equations of state, and modeled 
the decay of the magnetic field in two limiting cases, imposing at the
crust--core boundary either complete field expulsion by the superconducting 
core, or advection and freezing in a very highly conducting transition shell. 
The main result lies in the establishment of the existence of a tail in the 
rotation period distribution extending well beyond the shortest period observed 
so far ($P = 1.558$~ms), with only moderate dependence on the field suppression 
mechanism. For the softest equation of state the period distribution is still 
increasing at the shortest value before the onset of mass
shedding, where Possenti \etal\/ stopped their computations, while for the
stiffest one the period distribution had a wide maximum around $P = 2-4$~ms,
and a tail extending below this value. The fraction of objects with $P <
1.558$~ms is $\approx 1\%$ and $\approx 10\%$ for the stiff
and soft equation of state, respectively, all 
ending up with very small magnetic fields, $ \la 10^8$~G.

Though the accreted mass is larger when account is taken of the need to suppress
the magnetic field than when the magnetic field is neglected, the difference is
not very large (Burderi \etal, 1999). So, in order to appraise whether the
neutron stars
thusly formed may be (or not) supramassive, we simply consider Table I, from
Cook, Shapiro and Teukolsky (1994b). It shows that the total amount of mass
that needs being accreted from a companion in order to reach the supramassive
stage at the initial point of mass shedding depends strongly upon the equation
of state. For equation of state C, the neutron star collapses to a black hole 
even before reaching mass shedding. The soft equations of state, EoSs, (A, D, 
E, KC) have become supramassive; the intermediate EoSs (C, M, UT, FPS) are 
within $< 0.1$~M$_\odot$ of doing so, and will cross the threshold if accretion 
continues after the mass shedding point is reached (see below). The total 
amounts of mass required to become supramassive ($\approx 4/3$ of the 
difference in mass at infinity, Phinney and Kulkarni 1994, corresponding to a 
further $\approx 0.5$~M$_\odot$) are so modest that it seems likely that even 
the models based upon EoSs AU and UU will reach this stage, provided a donor of
sufficient mass is found: this too will be discussed below. EoSs L and
N$^\star$ are hopeless: the total amount of mass to be accreted corresponds to
$\approx 2$~M$_\odot$. If either of these EoSs were correct, there would be
no way to form SMNSs via accretion from a companion in a LMXB.

Both Cook, Shapiro and Teukolsky (1994b) and Possenti \etal (1999) halted
their computations when the mass shedding rotation rate is
reached, but there is nothing magic about this moment. Instead, 
Popham and Narayan (1991) and Paczynski (1991) argued that accretion
continues unimpeded, in Newtonian stars plus disks configurations, with
stars remaining close to the breakup angular speed, while mass and total
angular momentum increase. Other reasonable possibilities may contribute to
prolong mass accretion: the reduction of angular momentum through gravitational
wave losses (propitiated by the growth of a small stellar eccentricity)
or the setup of a spiral shock wave reducing the angular momentum
of incoming disk material, with the outward transport of angular momentum.
In any case, several avenues are possible which
would keep the neutron star marginally inside the mass shedding limit.
For this reason, it seems nearly certain that the intermediate
EoSs (C, M, UT, FPS) which are only $\approx 0.1$~M$_\odot$ away from being
supramassive, will reach this stage as mass accretion continues.

Since all soft and intermediate equations of state only require
$\approx 0.5$~M$_\odot$
to become supramassive, their companion star may be any star with mass
$\la 1$~M$_\odot$, exactly as discussed in the normal recycling model for
MSPs. EoSs AU and UT, instead,  require in total $\approx 1-1.1$~M$_\odot$ to
become supramassive. At first sight, this requirement might seem unsurmountable:
thermal stability in the mass exchange process through Roche lobe overflow
(Webbink \etal 1983) requires that the companion of the
NS has a mass below $5/6$ of the
neutron star's, \ie, $1.17$~M$_\odot$ for an initial
NS mass of $1.4$~M$_\odot$. Since the smallest He--core that may be left behind
is that of a star which took all Hubble time to evolve off the main sequence,
$0.16$~M$_\odot$, this leaves a maximum transferable mass of $1.01$~M$_\odot$,
less than required for either AU or UT. However, this is incorrect: the famous
requirement of $5/6$ths only occurs because Webbink \etal (1983) considered
Paczynski's (1967) approximation for the Roche lobe radius: using instead
Eggleton's (1983) formula, this requirement disappears. Webbink \etal (1983,
Eq. 15) show that the mass transfer rate is $\dot{M}_1 \propto
-1/(d\ln R_L/d\ln M_1)$ where
$R_L$ is the Roche lobe radius, and they argue that thermal stability in
the process requires $X \equiv d\ln R_L/d\ln M_1 > 0$. Using Eggleton's formula
$R_L/a = 0.49/(0.6+\ln(1+q^{1/3})/q^{2/3})$ where $a$ is the distance between
the two stars and $q\equiv M_1/M_2$ is the mass ratio, we find, under the
hypothesis of conservative mass transfer,
\begin{equation}
X = \frac{(1+q)(2(1+q^{1/3})\ln(1+q^{1/3})-q^{1/3})}{(1+q^{1/3})(1.8q^{2/3}+
3\ln(1+q^{1/3}))} > 0
\end{equation}
for every $q$! Also, dynamical stability
exists provided the donor mass is $ < 2$~M$_\odot$ (Rappaport \etal, 1995).
So we may consider as a possible companion for neutron stars
with EoSs AU and UT,
sub/giants of mass $\la 2$~M$_\odot$, where mass transfer is pushed forth by
donor radius expansion, in complete analogy with the model of Webbink \etal,
1983, except for donor mass. From Fig. 8b of Verbunt (1993), we see that a
giant or subgiant of nearly solar metallicity, of, say, $1.7$~M$_\odot$ manages 
to transfer at sub--Eddington rates $\approx 1.4$~M$_\odot$
to the neutron star, provided mass transfer begins when the giant core is
$\approx 0.2$~M$_\odot$; mass transfer will then leave behind a small
($\approx 0.3$~M$_\odot$), nearly inert He nucleus, with final period in the
range of $0.3$~d. We thus see that these systems provide attractive progenitors
for supramassive neutron stars, even in the case in which the applicable
EoS is either AU or UT, provided of course mass accretion is close to
conservative, an implicit assumption we made throughout, and that  mass  
can be accreted in sufficient quantities. 

Recent studies of binary pulsar masses (Thorsett and Chakrabarty 1999) seem to 
argue against significant mass accretion, but it should be noticed that, by 
investigating millisecond pulsars with periods exceeding $\approx 2\; ms$, the 
authors are investigating objects for which we know {\it a priori} that little 
mass need have been accreted, since their periods are long compared with SMNSs'.
We may expect different results when pulsars are chosen otherwise: a 
recent redetermination of the mass of Cyg X-2 finds $M = (1.8\pm 0.2)\; 
M_\odot$ (Orosz and Kuulkers 1999), departing from the narrow range of 
Thorsett and Chakrabarty.

The lowest magnetic field for the formation of
supramassive neutron stars may be lower than the empyrical value 
($\approx
2\times 10^8$~G) mentioned above (Possenti \etal\/ (1999) because 
when mass accretion from the companion begins to taper
off, or alternatively if mass accretion is intermittent, the Alfv\'en radius
(which scales as $\dot{M}^{-2/7}$) may expand further than the corotation
radius: the neutron star then goes through a new propeller phase
which slows its rotation. The overall effect is not large so that we shall 
consider in the following a maximum magnetic field
$q\times 10^8$~G, with $q \la 1$. 

\section{Further evolution}

Supramassive neutron stars are unstable to collapse to a black hole when angular
momentum losses reduce the initial angular momentum to about half of the
initial value; furthermore, these stars are peculiar in that evolution at
constant baryon number, but decreasing total angular momentum, makes them spin
up, rather than down; all of this is especially evident in Fig. 7-10-13-16 of
Salgado \etal, 1994.  Magnetic dipole losses cause a net torque which spins
down the neutron star in a time (Vietri and Stella 1998) 
$t_{sd} = 5\times 10^9\; yr (10^8\; G/B)^2$.
This time--scale is not strongly 
dependent upon EoS, but depends strongly upon whether the model is only 
marginally supramassive, or close to the absolute maximum mass (rotating or
not) for the given EoS, so that it may be considerably shorter under many
circumstances. Thus, a time $t_{sd}$ after becoming supramassive, the neutron 
star will collapse to a black hole. This time is reasonably long
when compared with typical mass accretion time--scales, which, as discussed
above, are typically determined by sub/giant nuclear evolution timescales. Thus
mass transfer will have long since ceased, and the immediate SMNS surroundings
will be reasonably baryon free. The companion star, in the meantime, will
have settled down as a low--luminosity, low--mass white dwarf, which is
not expected to pollute the environment either. Furthermore, we can gauge the
baryon--cleanliness of the SMNS surroundings at large if we assume that MSPs
are born through the same chain of events, except less extreme, for then we
know the Galactic distribution of MSPs. These are often located well outside the
Galactic disk, within an ISM with typical densities well in defect of
$n =  1$~cm$^{-3}$, which makes the total baryon mass within, say, $0.1$~pc, 
less than $10^{-5}$~M$_\odot$, more than enough to guarantee contact with the 
fireball model. We thus see that also this version of the formation scenario
guarantees a baryon clean environment, exactly like the different scenario
of Paper I.

The situation is clean even in the case in which the collapse 
occurs while mass transfer is still taking place. The total amount of baryons 
in the accretion disk is negligible: the disk crossing time is of order of 
$\approx 1\;$ month, which, with mass transfer rates $10^{-9}-10^{-8}\;M_\odot
\;yr^{-1}$, corresponds to much less than the maximum contamination value. The 
total amount of outlying mass from a wind is also rather small: for $\dot{M}_w 
\approx 10^{-9} M_\odot\; yr^{-1}$ and $v_w \approx 30\; km\; s^{-1}$, the total
mass within, say, $0.1\; pc$ is $3\times 10^{-6} M_\odot$, again negligible. 
The highly relativistic ejecta and $\gamma$ rays from the burst will hit the 
companion and form
a shock way inside the star's photosphere, so that local dissipation of
the ejecta kinetic energy will lead to the companion's inflating on the (long!)
Kelvin--Helmholtz time--scale, and the non--thermal afterglow emission
will not be contaminated by the re--radiated thermal component. 

The mechanism for the energy release is the same as discussed in paper I: once
the neutron star is destabilized, the innermost regions will collapse promptly 
to a black hole, while the equatorial matter, which is close to centrifugal
equilibrium, will just contract a little bit and begin orbiting the newly
formed black hole. A necessary condition which needs to be met is that this
equatorial material lies outside the innermost stable orbit. This can be
checked from Table I of Cook, Shapiro and Teukolsky (1994b) who show that
neutron stars which have reached the mass shedding regime have equatorial radii 
larger than the innermost stable orbit (see their column $j$), independent of 
EoS.  In paper I, we estimated the
amount of matter left behind as $\approx 0.1$~M$_\odot$;
this configuration is identical to the one hypothesized in most current models
(M\`esz\`aros 1999), and the debris torus is massive enough to power any burst,
especially in the presence of a moderate amount of beaming.

We now discuss the rate at which spun--up SMNSs
collapse to black holes. Since the timescales
involved are a fair fraction of the age of the Universe, and since star
formation evolves strongly in the recent past (Madau \etal, 1996), we have to
consider cosmological evolution of the population. However, from Fig.1 of White
and Ghosh (1998), it can be seen that the population of MSPs is roughly constant
(within the accuracy of the present, order of magnitude estimates) over the
redshift range $0 < z \la 1$ for most assumptions. There are currently an
estimated $5\times 10^4$ MSPs in the disk of the Galaxy (Lorimer 1995, Phinney
and Kulkarni 1994); assuming that there are as many systems in the bulge,
that a fraction $\beta$ of these are SMNSs, and that the typical timescale for
collapse to black hole is given by $t_{sd}$, the expected rate of collapses
in the Milky Way is $r = 10^5 \beta/t_{sd}= \beta/(5\times10^4\;{\rm yr})$, 
which is to be compared with
the inferred rate of GRBs, 1 every $3\times 10^7$~yr in an $L_\star$
galaxy like the Milky Way. Scaling to $\beta = 0.05$, a value intermediate
between the extremes of the simulations of Possenti \etal, we find that the two
rates agree for a beaming fraction $\delta\!\Omega/4\pi \approx 0.2(\beta/0.05)
$; this is consistent with the idea that these explosions do not require
extreme beaming fractions , since the explosion need not wade its way through
a massive stellar envelope, but immediately breaks free into a baryon clean
environment.

This model makes an easily testable prediction, because the location of bursts 
inside their host galaxies is the same as that of LMXBs, which are distributed 
at distances from the Galactic plane $\bar{z} \approx 1$~kpc, most
likely arising from kick velocities at the time of neutron star
formation (van Paradijs and White 1995). A similar $z$ distribution is 
observed for MSPs, $\bar{z} \ga 0.7$~kpc, and moderate transverse speeds.
Thus we would expect GRBs to cluster
around galactic disks (contrary to the binary pulsar merger model, where at
least some $50\%$ of all GRBs should be uncorrelated with the original birth
galaxies), but should not correlate with star forming regions (except for
SMNSs which form directly during a SN event, as discussed in paper I),
contrary to all scenarios involving massive stars. Also, 
the redshift distribution of GRBs within this model should be flatter than
the star formation distribution (again contrary to hypernovae), because the
redshift distribution of the MSP population is rather flat (White and Ghosh 
1998).

%It should also be pointed out that the explosions in this new formation scenario
%will not be surrounded by any large amount of matter. This stands in contrast
%with the scenario of paper I, where the ejecta produced in the SN explosion
%from which the SMNS formed directly, had time to move out only to $\approx
%10^{17} \; cm$ before the GRB.  This implies that the explanations of the 
%X--ray reburst and iron line from GRB 970508 and GRB 970828 (Vietri \etal, 
%1999, Lazzati \etal, 1999) will work only for the first formation scenario, in 
%qualitative agreement with the relative paucity of these events.

%In short, we have proposed here a new formation scenario for SMNSs, beyond the 
%traditional one of paper I: what we already know about magnetic
%field suppression in mass--transfering binary systems with a neutron star
%implies that $B$ can be damped enough to make the neutron star
%acquire mass and angular momentum and transform it into a SMNS, for any EoS
%except the hardest ever proposed. The triggering of the explosion is then a
%quintessentially general--relativistic effect,
%with the implosion/explosion occurring far from the disk, in a baryon--free
%environment. Event rates in agreement with observations can easily be
%obtained. Predictions about the distributions of GRBs' redshifts and
%locations within host galaxies clearly distinct from other models make this
%idea eminently testable.

We acknowledge helpful discussions with G. Ghisellini, A. Possenti 
and L. Burderi.

{}

\begin{table}
\begin{tabular}{|c|c|c|c|c|c|c|c|}
\multicolumn{8}{c}{Accretion from Keplerian disk onto a $1.4$~M$_\odot$
neutron star} \\
\hline
EoS$^a$ & $M_f/M_\odot^b$ & $M_i/M_\odot^c$ & $\bigtriangleup M/M_\odot^d$
&
$P_f^e$ & $R_i^f$ & $h_i^g$ & $M_{sm}/M_\odot^h$ \\
\hline
A & 1.77 & 1.57 & 0.428 & 0.604 & 9.59 & 2.82 & 1.66 \\
C & 1.74 & 1.54 & 0.389 & 0.894 & 12.1 & 0.27 & 1.86 \\
D & 1.76 & 1.56 & 0.405 & 0.730 & 10.7 & 1.70 & 1.65 \\
E & 1.76 & 1.57 & 0.414 & 0.656 & 10.0 & 2.38 & 1.75 \\
F & 1.52 & 1.59 & 0.172 & 0.715 & 9.21 & 3.20 & 1.46 \\
L & 1.80 & 1.52 & 0.443 & 1.250 & 15.0 & 0.00 & 2.70 \\
M & 1.74 & 1.50 & 0.367 & 1.490 & 16.7 & 0.00 & 1.80 \\
N$^\star$ & 1.84 & 1.53 & 0.484 & 1.080 & 13.6 & 0.00 & 2.64 \\
KC& 1.74 & 1.55 & 0.385 & 0.888 & 12.1 & 0.31 & 1.49 \\
AU& 1.79 & 1.58 & 0.446 & 0.701 & 10.4 & 2.01 & 2.13 \\
UU& 1.78 & 1.56 & 0.436 & 0.784 & 11.2 & 1.26 & 2.20 \\
UT& 1.78 & 1.57 & 0.429 & 0.754 & 10.9 & 1.52 & 1.84 \\
FPS& 1.76 & 1.56 & 0.416 & 0.747 & 10.9 & 1.56 & 1.80 \\
\hline
\end{tabular}
\caption{From Cook, Shapiro and Teukolsky, 1994b. $^a$ Equation of state;
$^b$ final total mass--energy: $^c$ initial rest mass; $^d$ accreted rest
mass;
$^e$ rotation period in $ms$; $^f$ initial circumferential radius in km;
$^g$ initial circumferential height of corotating marginally stable orbit in
km; $^h$ maximum static total mass--energy for EoS.}
\end{table}


\begin{references}
\reference{} Arons, J., 1993, \apj, 408, 160.
\reference{} Burderi, L., Possenti, A., Colpi, M., Di Salvo, T.,
D'Amico, N., 1999, \apj, 519, 285. 
\reference{} Camilo, F., Nice, D.J., Taylor, J.H., 1996, \apj, 461, 812. 
\reference{} Cook, G.B., Shapiro, S.L., Teukolsky, S.A., 1994a, \apj, 424,
823.
\reference{} Cook, G.B., Shapiro, S.L., Teukolsky, S.A., 1994b, \apjl, 423,
L117.
\reference{} Eggleton, P.P., 1983, \apj, 268, 368.
\reference{} Ghosh, P., Lamb, F.K., 1978, \apj, 223, L83.
\reference{} Ghosh, P., Lamb, F.K., 1992, in X--ray binaries and formation of
of binary and millisecond pulsars, ed. E. van den Heuvel and S. Rappaport,
Dordrecht, Kluwer (1992), 487.
\reference{} Illarionov, A., Sunyaev, R., 1975, \aap, 39, 185.
%\reference{} Lazzati, D., Campana, S., Ghisellini, G., \mnras, 304, L31.
\reference{} Lorimer, D.R., 1995, M\mnras, 274, L300.
\reference{} Madau, P., Ferguson, H.C., Dickinson, M.E., Giavalisco, M.,
Steidel, C.C., Fruchter, A., 1996, \mnras, 283, 1388.
\reference{} M\`esz\`aros, P., 1999, astro-ph. 9904038.
\reference{} Miller, M.C., Lamb, F.K., Psaltis, D., 1998, \apj, 508, 791.
\reference{} Orosz, J.A., Kuulkers, E., 1999, \mnras, 305, 132. 
\reference{} Paczynski, B., 1967, Acta Astron., 17, 287.
\reference{} Paczynski, B., 1991, \apj, 370, 593.
\reference{} Phinney, E.S., Kulkarni, S., 1994, \araa, 32, 591.
\reference{} Popham, R., Narayan, R., 1991, \apj, 370, 614.
\reference{} Possenti, A., Colpi, M., Geppert, U., Burderi, L., D'Amico,
N., 1999, \apjs, in press, astro-ph 9907231.
\reference{} Psaltis, D., Chakrabarty, D., 1999, \apj, 521, 332. 
\reference{} Rappaport, S., Podsiadlowski, P., Joss, P.C., Di Stefano,
R., Han, Z., 1995, \mnras, 273, 731.
\reference{} Rees, M.J., M\'esz\'aros, P., 1992, \mnras, 258, 41P.
\reference{} Ruderman, M., Zhu, T., Chen, K., 1998, \apj, 492, 267.
\reference{} Salgado, M., Bonazzola, S., Gourgoulhon, E., Haensel, P.,
1994, \aap, 291, 155.
\reference{} Stella, L., Vietri, M., 1999, \prl, 82, 17.
\reference{} Stella, L., Vietri, M., Morsink, S., 1999, \apjl, in press, 
astro--ph 9907346. 
\reference{} Thorsett, S.E., Chakrabarty, D., \apj, 512, 288.
\reference{} van der Klis, M., 1998, in Proccedings of the 3rd William
Fairbank Meeting, to appear, astro-ph 9812395.
\reference{} van Paradijs, J., White, N., 1995, \apjl, 447, L33.
\reference{} Verbunt, F., 1993, \araa, 31, 93
\reference{} Vietri, M., Stella, L., 1998, \apjl, 507, L45.
%\reference{} Vietri, M., Perola, G.C., Piro, L., Stella, L., 1999, \mnras, in
%press, astro--ph. 9906288.
\reference{} Webbink, R.F., Rappaport, S., Savonije, G.J., 1983, \apj, 270, 678.
\reference{} White, N.E., Ghosh, P., 1998, \apjl, 504, L31.


\end{references}
\end{document}